\documentclass[conference]{IEEEtran}
\usepackage{amsmath}
\usepackage [pdftex] {graphicx}
\usepackage{epstopdf}

\ifCLASSINFOpdf
\else
\fi
\begin{document}
%
\title{Mitigating Hardware Cyber-Security Risks in Error Correcting Decoders}

\author{\IEEEauthorblockN{Saied Hemati, \textit{Senior Member}, \textit{IEEE}}
\IEEEauthorblockA{Department of Electrical and Computer Engineering\\
University of Idaho\\
Moscow, Idaho, USA 83844-1023\\
Email: shemati@uidaho.edu\\
Telephone: (208) 885-6591
Fax: (208) 885-7579}
}

%


\maketitle

\begin{abstract}

This paper investigates hardware cyber-security risks associated with channel decoders, which are commonly acquired as a black box in semiconductor industry. It is shown that channel decoders are potentially attractive targets for hardware cyber-security attacks and can be easily embedded with malicious blocks. Several attack scenarios are considered in this work and suitable methods for mitigating the risks are proposed. These methods are based on randomizing the inputs of the channel decoder to obstruct the communications between attackers and the malicious blocks, ideally without changing the decoding performance.

\end{abstract}



%
\IEEEpeerreviewmaketitle

\section{Introduction}
Many companies that design cyber infrastructure, communications devices,  and  integrated circuits (ICs) use intellectual property (IP) elements developed by smaller and highly specialized companies, often without knowing exactly if the acquired IPs will do anything (possibly nefarious) beyond what they are supposed to do. These acquisitions are well justified financially and significantly reduce the cost associated with developing new ICs and products. This is also a healthy practice when different and unconventional specialties are needed. Nevertheless, security should not be compromised and special tools and design methodologies have to be developed to prevent any malicious activity carried out by the cyber systems that contains these ICs. 

A malicious block (also called a hardware Trojan) in a complex IC can be the result of an independent work of a few people in a large design team in a trustworthy company, or it can be hidden intentionally in an IP obtained in a supply chain or a commercial-off-the-shelf (COTS) product, or it can be an unintentional weakness or back-door that could be exploited by attackers. The malicious block could have been added in design, fabrication, packaging, testing, and assembling stages and it is non-trivial to discover the introduced blocks \cite{dod}-\cite{zz15}.

There have been some attempts for discovering hardware trustworthiness through screening the die's image to find any modifications or additional transistors. This is not an easy task as the number of transistors is huge and often dummy transistors and blocks are used for improving matching and manufacturability. Furthermore, processing the image of a die, sandwiched in a 3-D packaging would be challenging. Alternatively, generating test signals and characterizing a block by its input/output relationship, or adding a signature can only work when the number of inputs and outputs are small. Even restricting the purchases to a number of trustworthy providers cannot fix the security issue, since as mentioned earlier, a small number of designers can always add a few gates without their supervisor's permission or attention \cite{zz0}-\cite{zz15}.

The fact that many big companies are hacked every year indicates security issues have to be dealt with more rigorously \cite{ny}. While most cyber-attacks target stealing confidential or precious information, they often cannot harm the hardware, though they may temporarily interrupt a service. Hardware oriented attacks pose different challenges and can possibly cause permanent damages to devices and systems that we are relying on in our daily life and lack of them could jeopardize people's safety and cause economic damages \cite{zz0}.

This paper focuses on developing a design for assurance by mitigating  trustworthiness risks associated with cyber infrastructure hardware, more specifically in channel error-correcting decoders, which are an essential part of cyber systems that contain any digital communications systems or any computational system that uses memory. Channel decoders are important blocks in hardware trustworthiness analysis for the following reasons
\begin{enumerate} 
\item advanced channel decoders are complex circuits with millions of gates, thousands of floating point processing modules communicating through tens of thousands of wire connections. Yet it is possible to convert a decoder to a malicious block by using a few gates, which probably cannot be discovered by processing the die's image,
\item channel decoders directly interface with the outside world, which makes them an ideal block for receiving commands to start the nefarious act that could easily cause hardware or software failure,
\item channel decoders process noisy information, either caused by communication channel imperfections or by imperfections in storage media, and even a functional decoder cannot succeed all the time. A malicious channel decoder can easily claim false failures to block reception or retrieving stored information,
\item it is literally impossible to identify a malicious activity by running a number of test cases on resource limited simulation or emulation environments of a channel decoder as the number of inputs and outputs combinations are huge. A code of 2048 bits block length is not considered a long code \cite{ldpccode} and yet the number of different inputs is $2^{2048}$ in a hard-decision decoder and it is $2^{10,240}$ if soft-decoding with 5-bit resolution is used. It is thus impossible to verify all cases, and 
\item By developing suitable design methodologies and techniques to mitigate trustworthiness risks in channel decoders, which pose unique challenges, it becomes possible to mitigate trustworthiness risks in many other blocks and circuits that have less verification and implementation complexities, resulting in a design for assurance. 
\end{enumerate}

In this paper,
different attack scenarios will be investigated and possible remedies will be developed to mitigate the risks. The problem will be tackled by exploiting stochastic techniques to encrypt information internally within an integrated circuit and manipulating data transmission in its communication channels to avoid any unauthorized operation and isolate suspicious blocks. We assume that the cyber-security threats have not been found by screening the die or other techniques (\cite{zz0}-\cite{zz15}). This work develops general techniques to mitigate the trustworthiness risks instead of trying to eliminate hard-to-find malicious blocks in a channel decoder with arbitrary codes and block lengths.


  
\section{Cyber-security in Integrated Circuits}
Tackling cyber-security issues in integrated circuits and electronic systems is a relatively new problem and exhibits new challenges \cite{dod}-\cite{zz15}. A malicious block can be added to otherwise functional IP blocks by
\begin{enumerate} 
\item a few people in a design team within a trustworthy company acting without their supervisors' notice, 
\item companies with ill intentions selling IPs, possibly underpriced, to get into supply chain of trustworthy companies, and 
\item an honest mistake of the design team leaving a vulnerability or a back-door that can be later exploited by hackers. 
\end{enumerate}

In addition to design stage, a malicious block can be added during fabrication in a foundry, or during packaging, testing, assembling, and installing process. Again a few people in trustworthy companies can alter the original design without their supervisors' notice. Thus, there is always some chance that circuits and systems, even those developed by trustworthy companies would contain malicious blocks. This likelihood will be higher when there is no control over or access to companies selling their products in a supply chain or make commercial-off-the-shelf products.   

The malicious block could have been designed 

\begin{enumerate} 
\item to collect confidential information and pass it to unauthorized people (i.e., for espionage), 
\item to partially degrade the performance metrics (such as bit-error-rate, dynamic range, signal-to-noise ratio, lifespan, or energy-efficiency) in a competing product to win the market, 
\item to temporarily interrupt a service, make a system unstable, or cause malfunction, and 
\item to sabotage and permanently destroy the integrated circuit or electronic system (i.e., hardware attacks during cyber warfare). 
\end{enumerate}
These acts could be for political, economic, or military advantages. The first three items are commonly seen in cyber-attacks; however, software-based cyber-attacks cannot directly destroy an integrated circuit or electronics hardware. However, a malicious embedded circuit block can destroy an integrated circuit using many different techniques. For example by 

\begin{enumerate} 
\item intentionally causing latch-up problem (\cite{rabaey03}-\cite{johns}) in an integrated circuit by forward-biasing the substrate junctions to burn the integrated circuit and electronics system, 
\item short-circuiting the supply voltage or clock signal to damage power or clock tree, 
\item causing breakdown in gate oxide of MOS transistor by generating high-voltages using hidden capacitive voltage multipliers \cite{johns}, and 
\item increasing power consumption beyond thermal dissipation capabilities of the substrate to burn the integrated circuit and the electronics system. 
\end{enumerate} 
   
The malicious circuit can also generate noise to obstruct communications or normal operations or stop the related IP to induce malfunctions on integrated circuit or electronics systems. 
The malicious circuit can be most dangerous if they are activated simultaneously in a wide area, for example all smartphones or computers stop working at a specific moment based on 
receiving a direct command or a pattern embedded in the input signals or  
reaching to a preset time and date. 
Furthermore, a malicious circuit can be activated based on an internal clock or by exploiting parameters that show aging process in an integrated circuit such as electromigration \cite{johns}.

In order to protect integrated circuits, it is currently suggested to  (see for example  \cite{dod}-\cite{zz15})
\begin{enumerate} \itemsep1pt \parskip0pt \parsep0pt
\item limit purchases to trustworthy companies, 
\item process die's image to discover any alteration or any suspicious circuits, 
\item investigate input/output relationship to discover any abnormality, and
\item add signatures, develop physical unclonable function (PUF) modules, or use software-inspired Proof-
Carrying Code (PCC).
\end{enumerate}

\section{Deterministic Systems versus Stochastic Systems}

  The Achilles' heel of security in electronic and computer systems is, arguably, their universal architecture and deterministic behavior, which make them stationary targets \cite{pres}. This means that no individual identity is given to these systems and studying or tampering a single sample is enough to discover how similar models are working. In fact, we are making electronic and computer systems to be exactly identical. This feature is mainly the result of mass production of these systems and a desire to repair and possibly upgrade parts of the system in the future. While, this might seem a necessity for all similar systems, it is not exactly true for biological systems. Human beings do not think or observe exactly the same way, neither do they respond to external stimulates, diseases, and medications. Transplanting organs is not an easy task in human beings and it is quite tricky to deceive body's defense system not to reject a life-saving transplanted organ. We have always been envying machines that can last forever by replacing their faulty parts. However, this feature can also generate security vulnerability that malicious parts can be added to systems for nefarious purposes. 

A possible alternative to this paradigm is to promote stochastic systems, changing stationary targets to moving targets \cite{pres}, where each system follows a different trajectory in its normal operation and performs the required task differently to that extent that systems won't be exactly identical and even detailed information about the sample system does not provide the required knowledge to hackers to attack other similar systems.

One approach to mitigate malicious blocks, which are activated by a command message is to wrap deterministic parts in stochastic envelopes  or shields to make all communications within a system encrypted and hidden to external observers. In other words, building blocks of the system will operate in complete darkness and only process altered information using an encryption system that will be unique to each sample of an integrated circuit or system. 

Figure 1 shows an example for a general case when multiple modules are used in an electronic system. A seed is generated using a random noise that is sampled and quantized using an analog-to-digital converter (ADC) at the first boot-up. This seed will be a unique identity of the system and is sent to all modules within the system at the first boot-up and will be stored permanently, to form an encryption system that is device dependent. As an example, the random seed can be a long random binary sequence $\mathcal{B}$ with variable length, which is XORed with the output of each module ($X_{out}$) to generate an encrypted version of the output ($X_{out}^{encrypted}=\mathcal{B} \oplus X_{out}$). The module that receives the encrypted signal can only obtain  $X_{out}$ if it has $\mathcal{B}$ since $X_{out}=\mathcal{B} \oplus X_{out}^{encrypted}$. In this way, replacing the genuine modules with malicious modules will fail the system (as a rejected transplanted organ in human body) and is not possible. Also by tapping the wire connections, only the encrypted signals can be obtained, which will seem random (since it has been generated using random noise). Furthermore, even by successfully hacking a system only $\mathcal{B}$ can be obtained but it does not reveal any information regarding other samples of the electronic system.

\begin{figure}
	\centering
 	\includegraphics[width=0.47\textwidth]{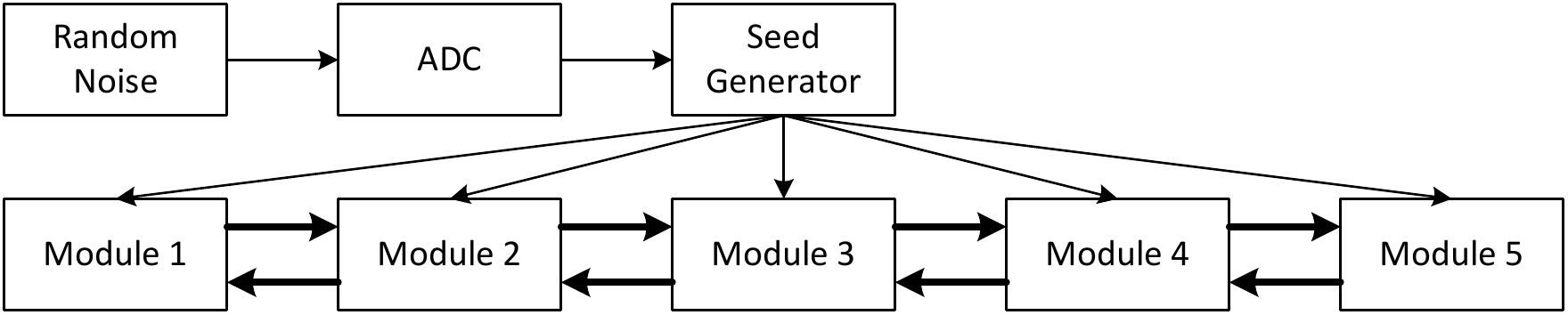} 
	\caption{\normalsize {Generating a random seed in the first boot-up to initialize encryption in modules of an electronic system. Any attempt for replacing genuine modules with malicious modules or tapping the communications among modules will be difficult and sample dependent.}}
	\label{active}
\end{figure}

\section {Trustworthiness in Channel Decoders}
Channel coding is an indispensable part of any modern digital communications system. It works by introducing some redundancy to the transmitted information to make communications reliable as shown in Figure 1. At the receiver side, a decoder exploits the added redundancy to find and possibly correct any mistake that may have happened due to transmission channel imperfections \cite{shan} and \cite{shu}.

Let's assume a binary linear block code is used and an encoder, which is located in the transmitter, maps a message vector $m$ to a codeword vector $c$ using a generator matrix $\boldsymbol{G}$ and channel noise ($n$) is additive and the decoder receives $r$ vector which is equal to $c+n$. 

\begin{figure}
	\centering
 	\includegraphics[width=0.47\textwidth] {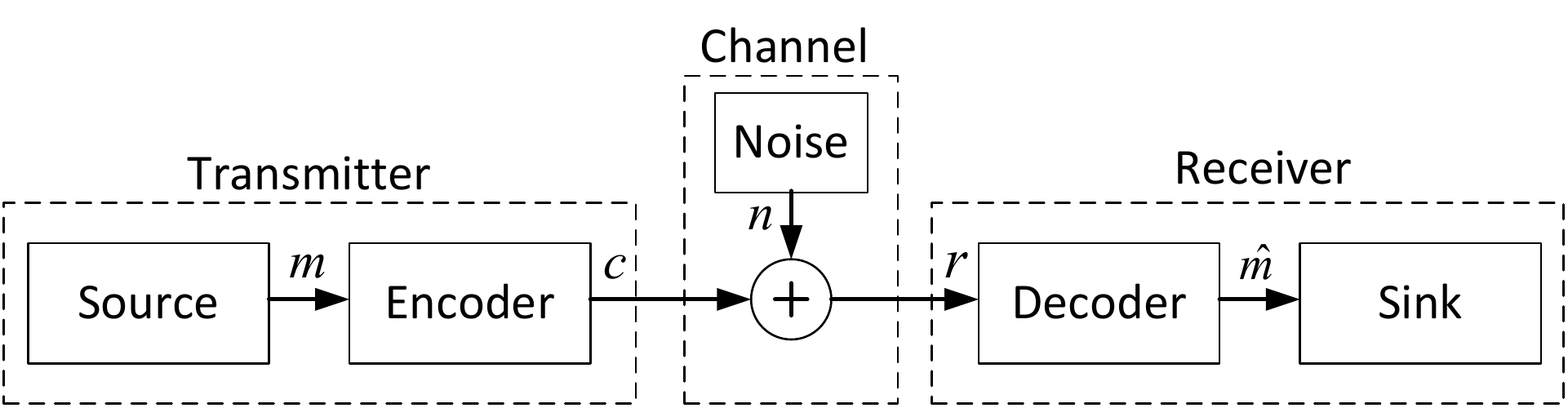}
	\caption{\normalsize {A simple model for digital communications systems.}}
	\label{active2}
\end{figure}

In the following sections, it is assumed that the attackers can broadcast signals that are received by a channel decoder embedded with a malicious block.

\subsection{Case-I: The Malicious Circuit is Activated by a Sequence of Codewords}
Let's assume the malicious circuit inside a channel decoder is activated, to say cause latch-up problem to burn the IC, if five codewords $c_1$, $c_2$, $c_3$, $c_4$, and $c_5$  (or their corresponding messages, i.e., $m_1$, $m_2$, $m_3$, $m_4$, and $m_5$) are detected in a specific sequence. Obviously, it is not feasible to generate all possible scenarios like this to discover the malicious circuit. 

Based on the following Lemma, a stochastic envelope can be developed to surround the channel decoder.

\textbf{Lemma}: If $\mathcal{D}$ is a maximum likelihood decoder or any suboptimal decoder that equally treats codewords and $\mathcal{D}(r)=\hat{m}$  ($\hat{m}=m$ if decoding is successful with no undetectable error) then  $\mathcal{D}(r+c_x)=\hat{m}+m_x$ where $m_x$ is a random message and $c_x$ is its corresponding codeword. In hard-decision decoding, $r+c_x=c \oplus e \oplus c_x$, where $e$ is the error vector. In soft-decision decoding, the polarity of log-likelihood ratio (LLR) representation \cite{shu}, corresponding to each bit in $r'=r+c_x$ will be equal to the binary addition of $c_x$ with the hard-decision corresponding to each bit in $r$. Obviously, the magnitude of LLR representation will not be changed.  
This Lemma can be proved based on the fact that for any linear block code, the addition of every two codewords is also a codeword.$\clubsuit$

Therefore, the received information ($r$) can be mapped randomly to another vector ($r'=r+c_x$) before it is applied to the decoder. The output of the decoder will be $\hat{m}+m_x$ and will be added again with $m_x$ to result the expected output of the decoding ($\hat{m}$) and the decoder cannot discover what was the codeword (see Figure 3). Meanwhile, $m_x$ is a randomly selected message that can be changed each time based on a different seed in different samples, making it difficult to tamper a sample device to attack other devices using the malicious decoder.
The probability of observing the sequence of $c_1$ , $c_2$, $c_3$, $c_4$, and $c_5$ for a device using the proposed stochastic system will not be exactly zero. However, the extremely unlikely event of activating the malicious block by chance (based on random $c_x$s)  will not happen when the attackers want, thus makes the orchestrated attack practically ineffective.  

It is also important to note the malicious circuit, in this example, can be very small (in the order of a few hundred gates), which can be easily hidden in advanced multi-million gate decoders \cite{hem0}-\cite{hem9}.

\subsection{Case-II: The Malicious Circuit is Activated by a Sequence of Error}
A more challenging case is when the attack command is embedded on a superficial error vector ($e_{sup}$) that renders shifting codewords useless.  By superficial error, we mean the attackers transmit a vector, which is not a codeword, i.e., $c \oplus e_{sup}$  instead of $c$ and overpower the natural channel noise in a way that the malicious block can extract $e_{sup}$ or some message embedded in it from the received information. It is interesting to note by shifting the codeword (Case-I),  the error vector will not change. This is obvious by noting that shifting codewords changes $c \oplus e$ to $c \oplus c_x \oplus e$, which does not have any impact on the error vector.
 
Deterministic techniques, arguably fail to mitigate this attack. However,  stochastic techniques such as stochastic Chase \cite{hem4} and dithered belief propagation \cite{hem5} decoding algorithms that intentionally add random noise to the received vector before decoding, to possibly improve the decoding performance and rely on several decoding trials, can be used for changing the noise vector and disabling this attack. However, there is no guarantee that the impact of $e_{sup}$ can be completely removed without adding too much noise and degrading the communications system. 

\begin{figure}
	\centering
 	\includegraphics[width=0.45\textwidth]{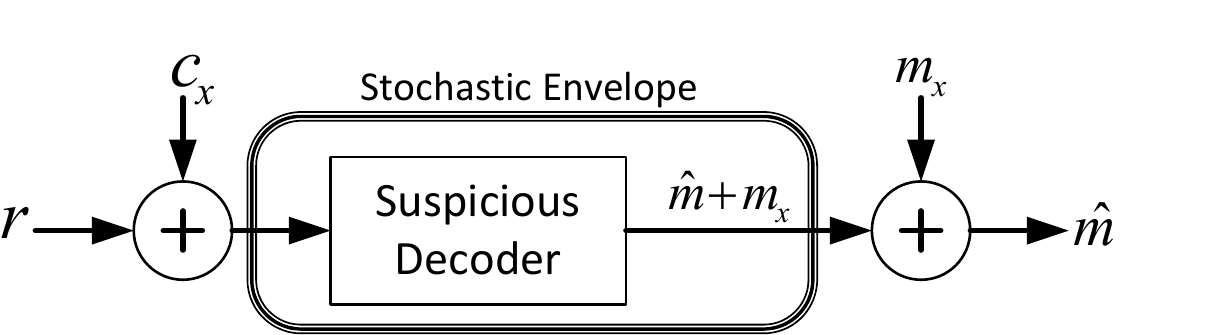}
	\caption{\normalsize {A suspicious decoder wrapped in a stochastic envelope cannot receive any command based on codewords. $m_x$ is a random message and $c_x$ is its corresponding codeword.}}
	\label{active3}
\end{figure}

\vspace{-0.3cm}
\subsection{Case-III: The Malicious Circuit is Activated by a Sequence of Failure and Success in Decoding}
In this case, instead of a sequence of codewords or a noise pattern, a sequence of failure and success in decoding triggers the malicious circuit. For example, if a failed decoding (when no $\hat{m}$ is output) is represented by \emph{f}, and a successful decoding (when $\hat{m}$ is output even if it is not equal to $m$) is represented by \emph{s}, then an example sequence could be \emph{fsfsffssfffsssffffssssfsfsffssfffsssffffssssfsfsffssfffsssffffssssfsfsffssfffsssffffssss}.

The sequence should be long enough to make the likelihood of naturally observing the sequence near zero. As an example, let's assume the chance of failure in decoding is $10^{-5}$, then the probability of observing the above sequence is almost $10^{-220}$. 

In this case the malicious circuit will be extremely simple and only consists of an 88 bit shift register and a small number of simple gates, which makes it even easier to hide in a large decoder circuit. Obviously, the techniques mentioned in previous examples do not help in this case, because they do not change the failure-success sequence. 

To mitigate this malicious block, it is possible to

\begin{enumerate} 
\item erase any memory inside the decoder by powering it off or by other means after each decoding operation. The idle time should be long enough to discharge any capacitive memory that might have been used, 

\item randomly reorder the received vectors using a stack at the input to prevent the decoder know the real sequence, and

\item use redundant decoders and distribute the received blocks randomly among the decoders, preventing the decoders know anything about the real sequence or its statistics.

\end{enumerate}

In Figure 4, redundant decoders are utilized to change the failure/success statistics and facilitate powering off the other decoders.

If internal memories cannot be erased or information is stored on capacitors hidden in the circuit, which retain the information during short power-off intervals (a few $\mu$s in low-latency communications systems), then randomizing the sequence of received vectors wont be completely effective. Unfortunately, utilizing many redundant decoders is not a viable approach because channel decoders are big circuits.

If the number of redundant decoders is small, attackers can utilize communication techniques to convey the attack command to the malicious channel decoder. For example,  they can use repetition coding scheme \cite{shu}, i.e., each failure or success is sent multiple time, for example 10 times. It means $f$ is replaced by \emph{ffffffffff} and \emph{s} is replaced by \emph{ssssssssss}. If the received vectors are sent randomly to one of two available channel decoder in Figure 4, and the malicious decoders are smart enough to track timing, the sequence will consist of a number of \emph{f} and a number of \emph{I} (standing for idle). For example, \emph{IfIfIIffII}, which can be easily recognized as an \emph{f} unless idle time intervals are replaced by completely random decoding. 

Nevertheless, if the number of random decoding is not large enough, attackers can still send their message to the malicious channel decoder \emph{by treating this randomizing process similar to a noisy communication channel and utilizing the decoding capability of the malicious decoder to recover the original sequence}. After all, a channel decoder is used to recover a message transmitted through a noisy channel.

\begin{figure}
	\centering
 	\includegraphics[width=0.45\textwidth] {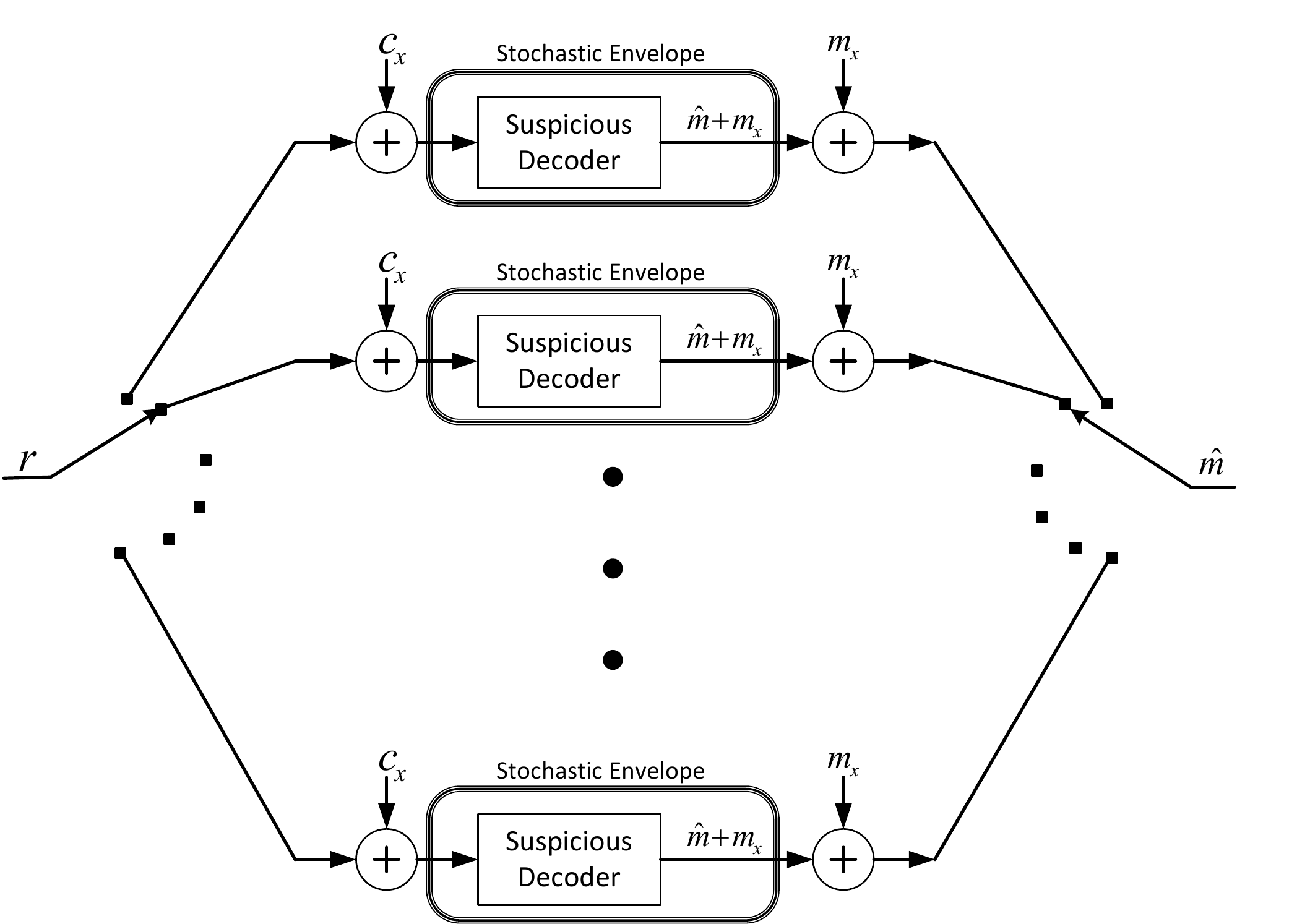}
	\caption{\normalsize {Using redundant decoders to prevent exploiting the sequence of success/failure or other statistical information.}}
	\label{active4}
\end{figure}

\section {Conclusion}
In this paper different scenarios were discussed for mitigating trustworthiness risks in channel decoders. 
It was demonstrated that by completely isolating a block in a circuit from the outside world, by wrapping it in a stochastic envelope in a way that no direct data or statistical information is passed to the block, many attacks can be mitigated. We used additional redundancy to randomize statistics and developed techniques for randomly shifting codewords. Stochastic techniques can also be utilized to further randomize inputs of a malicious decoder to hide data from the processing modules.

It was also highlighted that a malicious embedded block can destroy an integrated circuit and harm a cyber-hardware in a second, which makes it different and more dangerous than common software cyber-attacks. Furthermore, it was shown that a malicious block can be very small and can remain hidden in a large circuit. 

These observations and other preliminary results represent just the tip of an iceberg that cyber infrastructure hardware would face, and require further investigations and developments to quantify and mitigate their harmful effects. Developing viable solutions and approaches for safeguarding integrated circuits and electronic systems utilizing channel decoders necessitates these comprehensive investigations too. 

\end{document}